\documentclass[useAMS,usenatbib]{mn2e}

\usepackage{graphicx,epsfig,subfigure}
\usepackage{url}

\usepackage{amsmath}
\usepackage{amssymb}

\usepackage{threeparttable}
\usepackage{float}

\title[SPM Seeing Statistics]
{Astroclimate at San Pedro M\'artir I: 2004\,--\,2008 Seeing
Statistics from the TMT Site Testing Data\thanks{Based on
observations obtained at the Observatorio Astron\'omico Nacional
at San Pedro M\'artir, Baja California, M\'exico, operated by the
Instituto de Astronom\'{\i}a, Universidad Nacional Aut\'onoma de
M\'exico.}}

\author[S\'anchez, Cruz-Gonz\'alez et al.]
{L.~J.~S\'anchez,$^1$\thanks{E-mail: leonardo@astro.unam.mx}
I.~Cruz-Gonz\'alez,$^1$ J.~Echevarr\'{\i}a,$^1$
A.~Ruelas-Mayorga,$^1$
\newauthor
A.~M.~Garc\'{\i}a,$^1$ R.~Avila,$^2$ E.~Carrasco,$^3$
A.~Carrami\~nana,$^3$ and A.~Nigoche-Netro$^4$\\
$^1$Instituto de Astronom\'{\i}a, Universidad Nacional Aut\'onoma
de M\'exico, Cd. Universitaria, \\M\'exico, D.F. 04510, M\'exico\\
$^2$Centro de F{\'\i}sica Aplicada y Tecnolog{\'\i}a Avanzada,
Universidad Nacional Aut\'{o}noma de M\'{e}xico, \\
Santiago de Quer\'{e}taro, Qro. 76000, M\'exico \\
$^3$Instituto Nacional de Astrof{\'\i}sica, \'Optica y
Electr\'onica, Tonantzintla, Pue. 72840,  M\'exico\\
$^4$Instituto de Astronom{\'\i}a y Meteorolog{\'\i}a, Universidad
de Guadalajara, Guadalajara, Jal. 44130, M\'exico}

\begin{document}

\date{Accepted 2012. Received 2012 March; in original form 2012 March}

\pagerange{\pageref{firstpage}--\pageref{lastpage}}
\pubyear{2012}

\maketitle

\label{firstpage}


\begin{abstract}
We present comprehensive seeing statistics for the San Pedro
M\'artir site derived from the Thirty Meter Telescope site
selection data. The observations were obtained between 2004 and
2008 with a Differential Image Motion Monitor (DIMM) and a Multi
Aperture Scintillation Sensor (MASS) combined instrument
(MASS\,–-\,DIMM). The parameters that are statistically analised
here are: whole atmosphere seeing –measured by the DIMM–; free
atmosphere seeing –-measured by the MASS-–; and ground-layer
seeing (GL) –-difference between the total and free-atmosphere
seeing-–. We made a careful data coverage study along with
statistical distributions of simultaneous MASS\,–-\,DIMM seeing
measurements, in order to investigate the nightly, monthly,
seasonal, annual and global behaviour, as well as possible hourly
seeing trends. Although this campaign covers five years, the
sampling is uneven, being 2006 and 2007 the best sampled years in
terms of seasonal coverage. The overall results yield a median
seeing of 0.78 (DIMM), 0.37 (MASS) and 0.59\,arcsec (GL). The
strongest contribution to the whole atmosphere seeing comes,
therefore, from a strong ground layer. We find that the best
season is summer, while the worst one is winter, in accordance
with previous studies. It is worth noting that the best yearly
results are correlated with the best sampled years. The hourly
analysis shows that there is no statistically significant tendency
of seeing degradation towards dawn. The seeing values are slightly
larger than those reported before. This may be caused by climate
changes.
\end{abstract}




\section{Introduction}
\label{sec:intro}

In the early seventies a new observing site began operations at
the Sierra de San Pedro M\'artir (SPM), Baja California, M\'exico.
The site was selected through satellite photographs, and was found
to be one of the three best cloud-free areas in the world.

We now know that this site is one of the best astronomical
locations in the world and has been considered by the
international astronomical community as a potential place for
large telescopes to be built in the near future being a candidate
site of projects such as the Large Synoptic Survey Telescope
(LSST) and the Thirty Meter Telescope (TMT), as well as other
astronomical projects.

Several climatological properties have been reported, mainly
during the first years of operation of the Observatorio
Astron\'omico Nacional (OAN) at SPM
\citep{1971BOTT....6...95M,1973Mercu...2....9M,1972BOTT....6..215M,1977RMxAA...2...43A,
1982sham.conf..311A,1984ESOC...18....3W} and later by
\citet{1992RMxAA..24..179T,2003RMxAC..19...75T,
1998RMxAA..34...47E,2001RMxAA..37..213H,
2003RMxAC..19...99M,2003RMxAC..19...37M,2003RMxAC..19..103C,
2005PASP..117..104C,2006PASP..118..503A,2007RMxAC..28....9T,
2007RMxAC..31..113A,2008RMxAA..44..231B,2009RMxAA..45..161O,
2010RMxAA..46...89B,2011RMxAA..47..409A} and
\citet{2012MNRAS.420.1273C}.


\begin{table*}
\centering \caption{SPM seeing monitoring results (whole
atmosphere seeing).} \label{tab:ResultsSeeing}

\begin{tabular}{crrrr}
\hline Method &  1st Quart. & Median &
3rd Quart. & Nights \\
 & (arcsec)& (arcsec) & (arcsec)& \\
\hline

STT$^{1}$   &  0.50& 0.61 &  & 386 \\

CM$^{1}$    &  0.48& 0.63 &  & 99 \\

CM / STT$^{1,a}$ &  0.46 & 0.58 &  & 57 \\

DIMM$^{2}$ &  & 0.50;~0.75$^{b}$ & & 14  \\

DIMM$^{2}$ &  0.61 & 0.77 & 0.99 & 31  \\

DIMM$^{3}$ &  0.48 & 0.60 & 0.81 & 123 \\

MASS\,--\,DIMM$^{4}$ & 0.61 & 0.79 & 1.12 & $^{c}$ \\

SciDAR$^{5}$ &  0.50 & 0.68 & 0.97  & 27 \\

\hline

\end{tabular}

$^{1}$\,\citet{1998RMxAA..34...47E},~$^{2}$\,\citet{2002AA...396..723C},~$^{3}$\,\citet{2003RMxAC..19...37M},~$^{4}$\,\citet{2009PASP..121.1151S},~$^{5}$\,\citet{2011RMxAA..47...75A} \\
$^{a}$ CM simultaneous with STT,~$^{b}$ Bimodal distribution,~$^{c}$ (2004 Oct. -- 2008 Feb.) \\

\end{table*}


Results on the seeing and optical turbulence above the ground can
be found in
\citet{1998PASP..110.1106A,1998RMxAA..34...47E,2002AA...396..723C,
2007RMxAC..31...71A,2007RMxAC..31...93S}  and
\citet{2011RMxAA..47...75A}, as well as atmosphere modelling by
\citet{2001A&A...366..708M,2002A&A...382..378M,2003RMxAC..19...63M,2004RMxAA..40....3M}
and \citet{2004RMxAA..40...81V}. Extinction and opacity studies
have also been made by
\citet{2001RMxAA..37..187S,2003RMxAC..19...90H} and
\citet{2003RMxAC..19...81P}. Site prospection studies within the
SPM site have been made by \citet{2007RMxAC..31..122S} and
\citet{2008RMxAA..44..231B}. Comprehensive reviews on the site can
be found in \citet{2003RMxAC..19.....C,2004SPIE.5382..634C,
2007RMxAC..28....9T,2007RMxAC..31...47T} and
\citet{2007RMxAC..28....1W}.

In this article, we analyse the data collected by the TMT Project
Site Survey at the SPM site and discuss and compare results of
Differential Image Motion Monitor (DIMM) and a Multi Aperture
Scintillation Sensor (MASS) with previous studies. The
observations include data in the time period 2004 October to 2008
February analysed by \citet{2009PASP..121.1151S}, plus data taken
between 2008 February to 2008 August whose analysis has not been
presented anywhere in the astronomical literature, improving
considerably the analysis of the complete TMT MASS\,--\,DIMM
seeing survey at SPM and supplementing previous seeing results
published in the literature.

We also present in a detailed manner the number of observations
and the percentage of time covered by them. This a novel way of
treating seeing data.

In the spirit of providing the astronomical community with
detailed information of the seeing behaviour in SPM, we present
nightly, monthly, seasonal, annual and global statistics. We too,
present an analysis of the hourly seeing trend which yields
different results from those found by \citet{2009PASP..121.1151S}.

A summary of prior SPM seeing studies are mentioned in
Section~\ref{sec:spmsite}, SPM TMT site testing instrumentation
and data coverage are presented in Section~\ref{sec:tmtdata},
followed by the results and detailed statistics in
Section~\ref{sec:results}.


\section{The SPM Site: Seeing Studies}
\label{sec:spmsite}

The atmospheric turbulence is usually studied through a number of
parameters, one of which is known as seeing ($\varepsilon$). The
relation between $\varepsilon$, Fried parameter $r_{o}$, and the
turbulence integral is given by
\begin{equation}
\varepsilon   = 0.98\frac{\lambda}{{r_{o}}} = 5.25\lambda^
{-{1/5}}[{\int_0^\infty{C_N^2\left(h\right)dh}}]^{3/5},
\label{eq:seeingeq}
\end{equation}
where $\lambda$ is the wavelength and ${\int_0^\infty  {C_N^2
\left( h \right)dh}}$ is the optical turbulence energy profile
\citep{1981PrOpt..19..281R}.

We note that the seeing units are given in arcseconds (arcsec) and
since seeing is a wavelength-dependent turbulence parameter it is
usually calculated for 0.5\,$\mu$m. The seeing value is also
corrected for the direction of observation and is referred to zero
zenith angle.

Early works on local SPM seeing conditions were done by
\citet{1971BOTT....6...95M} and \citet{1971PASP...83..401W}, and
several recent studies have been carried out by
\citet{1998RMxAA..34...47E,2002AA...396..723C} and
\citet{2003RMxAC..19...37M}.  Optical turbulence studies have been
made by
\citet{1998PASP..110.1106A,2004PASP..116..682A,2006PASP..118..503A,2007RMxAC..31...71A}
and \citet{2011RMxAA..47...75A}. A comprehensive review of some of
these studies can be found in \citet{2003RMxAC..19...41E}.

The \citet{1998RMxAA..34...47E} observations were obtained with
two seeing monitors and a Micro Temperature Array tower (MTA); the
Site Testing Telescope (STT) from Steward Observatory was designed
to observe Polaris, while the Carnegie Monitor (CM) could observe
and track any star. The MTA consisted of platinum detectors, which
measured temperature differences located at different heights,
from 4 to 28\,m. These authors report a median seeing of
0.61\,arcsec with a first quartile of 0.50\,arcsec and a decrease
of 0.1\,arcsec at a height of 15\,m. The STT observations covered
386 nights, while the CM observations covered 114 nights, spanning
a three year period. A summary of their results is shown in
Table~\ref{tab:ResultsSeeing}.

\citet{2002AA...396..723C} measured the wavefront outer scale and
included ground-based seeing measurements with a DIMM monitor.
They reported observations during 31 nights. The telescope was
located at two sites: at the CM tower and at a low altitude
location. In their article they found a bimodal distribution
during their 2000 December campaign with peaks centred at 0.50 and
0.75\,arcsec, and an overall seeing median of 0.77\,arcsec
including all observations during the 31-night period.

\citet{2003RMxAC..19...37M} conducted a study with
the same DIMM monitor during 123 nights spanning almost a three
year period. The monitor was located at the CM tower at a height
of 8.3\,m. These authors reported a median of 0.62\,arcsec and a
first quartile of 0.49\,arcsec.

\citet{2009PASP..121.1151S} analyse data collected by the TMT
Project Site Survey at five different sites which include SPM. A
partial set (2004 Oct. to 2008 Feb.), of combined DIMM and MASS
data, yields seeing median values for the ground layer
0.58\,arcsec, for the free atmosphere 0.37\,arcsec and for the
whole atmosphere 0.79\,arcsec.

Concerning optical turbulence profiling,
\citet{1998PASP..110.1106A,2004PASP..116..682A} monitored the
vertical distribution of the optical turbulence strength using a
Generalized SciDAR (GS). Two observation campaigns have been
carried out at the OAN\,--\,SPM in 1997 and 2000. In 1997, the GS
was installed at the 1.5\,m and 2.1\,m telescopes (SPM1.5 and
SPM2.1) for 8 and 3 nights (March and April), respectively,
whereas in 2000 the instrument was installed for 9 and 7 nights
(May) at SPM1.5 and SPM2.1, respectively.

\citet{2011RMxAA..47...75A} performed a recalibration of the
${C_N^2 \left( h \right)}$ profiles obtained in the OAN\,--\,SPM,
following the results of a theoretical work on the normalization
procedure used in the GS data reduction method
\citep{2009OExpr..1710926A}. From a statistical analysis of the
recalibrated profiles, \citet{2011RMxAA..47...75A} have found that
the seeing in the first two kilometers, in the free atmosphere,
and in the whole atmosphere had median values of 0.38, 0.35 and
0.68\,arcsec, respectively.

A synthesis of the non-differential monitors (STT \& CM),
differential (DIMM) and SciDAR seeing monitoring campaigns is
shown in Table~\ref{tab:ResultsSeeing}.


\section{TMT Site Testing Data}
\label{sec:tmtdata}


\subsection{The 2004\,--\,2008 TMT site testing campaign}
\label{subsec:tmtcampaign}

SPM was one of the five candidate sites selected by the TMT site
testing team \citep{2009PASP..121..384S}. Three southern
hemisphere sites were studied, Cerros Tolar, Armazones and
Tolonchar in northern Chile, and two northern hemisphere sites,
the 13 North (13\,N) site on Mauna Kea, Hawaii in the United
States and SPM  in  Mexico. During a period of approximately five
years, from 2004 to 2008, the TMT group measured the atmospheric
properties of each site with the same instrumentation and at least
2.5 annual cycles of data were acquired on each of the candidate
sites. As is described in detail by \citet{2009PASP..121..384S} a
suite of eight instruments were deployed in the candidate sites.
The data acquisition methodology is also presented in their work.
The acquired data of all the instruments is kindly available to
the public at \url{http://sitedata.tmt.org/} by the TMT
organisation.


\begin{figure}
  \begin{center}
    \includegraphics[width=\columnwidth,keepaspectratio=true]
    {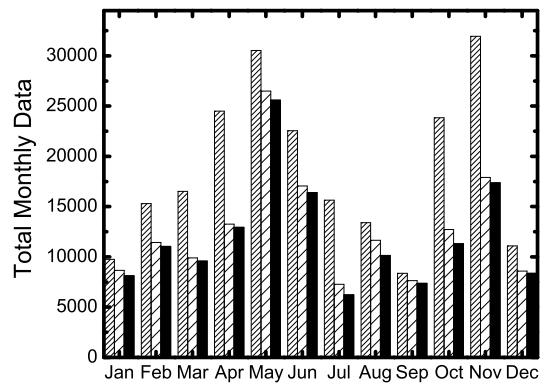}
  \end{center}
\caption{Total monthly data for the  2004\,--\,2008 period.
Patterns: DIMM (dense), MASS (sparse), simultaneous MASS\,--\,DIMM
(dark).}
  \label{fig:AllDataAmount_DIMM_MASS_Simul}
\end{figure}



\begin{table*}
\caption{Amount of data collected and coverage percentage for
DIMM.} \label{tab:AllDataCovDIMM} \centering
    \begin{tabular}{lrrrrrrrrrrrr}
        \hline
        \multicolumn{13}{c}{\bf{Year}}           \\
        \hline
    Month   & \multicolumn{2}{c}{2004} & \multicolumn{2}{c}{2005} & \multicolumn{2}{c}{2006}
            & \multicolumn{2}{c}{2007} &  \multicolumn{2}{c}{2008}
            & \multicolumn{2}{c}{\bf{Total}}\\
        \hline

    January     & -- & --   & 129 & 0.9\,\% & 3371 & 23.1\,\%   & 6263 & 43.1\,\%
                & 0 & 0.0\,\%   & 9763 & 16.8\,\%\\
    February    & -- & --   & 0  & 0.0\,\%     & 3082 & 23.5\,\%   & 7767 & 60.9\,\%
                & 4449 & 35.5\,\% & 15298 & 30.0\,\%\\
    March       & -- & --   & 0 & 0.0\,\%   & 5740 & 46.3\,\%   & 5992 & 47.0\,\%
                & 4794 & 38.9\,\%       & 16526 & 33.0\,\%\\

        \hline
    \bf{Winter}      & & & & & & & & & &
                & 41587 & 26.6\,\%\\
        \hline

    April       & -- & --   & 2846 & 25.5\,\% & 8185 & 74.4\,\% & 3901 & 36.4\,\%
                & 9557 & 86.7\,\%       & 24489 & 55.7\,\% \\
    May         & -- & --   & 6755 & 64.9\,\% & 7640 & 73.5\,\% & 8472 & 81.5\,\%
                & 7657 & 73.6\,\%       & 30524 & 73.4\,\% \\
    June        & -- & --   &   5740 & 59.9\,\% & 6609 & 69.0\,\% & 3959 & 41.4\,\%
                & 6233 & 65.1\,\% & 22541 & 58.8\,\% \\

        \hline
    \bf{Spring}      & & & & & & & & & &
                & 77554 & 62.6\,\%\\
        \hline

    July        & -- & --   & 628 & 6.3\,\% & 5064 & 50.3\,\% & 5792 & 57.7\,\%
                & 4155 & 41.2\,\% & 15639 & 38.9\,\%\\
    August      & -- & --   & 0 & 0.0\,\%   & 6578 & 60.3\,\% & 6174 & 56.1\,\%
                & 642 & 6.1\,\% & 13394 & 30.6\,\% \\
    September   & -- & --   & 0 & 0.0\,\%   & 8361 & 71.2\,\%  & 0 &
                0.0\,\% & -- & --    & 8361 & 23.8\,\% \\

        \hline
    \bf{Summer}      & & & & & & & & & &
                & 37394 & 31.1\,\%\\
        \hline

    October     & 6081 & 45.5\,\%   & 445 & 3.2\,\% & 8712 & 66.1\,\% & 8612 & 63.7\,\%
                & -- & -- & 23850 & 44.6\,\%\\
    November    & 3887 & 28.0\,\%   & 10134 & 73.6\,\% & 9916 & 72.1\,\% & 8004 & 57.7\,\%
                & -- & --   & 31941 & 57.8\,\%\\
    December    & 0 & 0.0\,\%   & 5165 & 35.1\,\%       & 5933 & 40.3\,\% & 0 &
                0.0\,\% & -- & --   & 11098 & 18.8\,\%\\

        \hline
    \bf{Autumn}      & & & & & & & & & &
                & 66889 & 40.4\,\%\\
        \hline

    \bf{Total}       & 9968 & 36.7\,\%  & 31842 & 22.4\,\% & 79191 & 55.8\,\% & 64936 & 45.4\,\%
                & 37487 & 43.4\,\%  &  223424 &  40.2\,\% \\
        \hline

\end{tabular}
\end{table*}


\begin{table*}
\caption{Amount of data collected and coverage percentage for
MASS.} \label{tab:AllDataCovMASS} \centering
\begin{tabular}{lrrrrrrrrrrrr}
        \hline
        \multicolumn{13}{c}{\bf{Year}}           \\
        \hline
    Month   & \multicolumn{2}{c}{2004} & \multicolumn{2}{c}{2005} & \multicolumn{2}{c}{2006}
            & \multicolumn{2}{c}{2007} &  \multicolumn{2}{c}{2008}
            & \multicolumn{2}{c}{\bf{Total}}\\
        \hline

    January     & -- & -- & 241 & 1.7\,\% & 2663 & 18.2\,\% & 5768 & 39.7\,\%
                & 0 & 0.0\,\% & 8672 & 14.9\,\%\\
    February    & -- & -- & 0  & 0.0\,\% & 2448 & 18.7\,\%   & 6451 & 50.6\,\%
                & 2536 & 20.3\,\% & 11435 & 22.4\,\%\\

    March       & -- & --   & 0 & 0.0\,\%   & 4669 & 37.7\,\%   & 4890 & 38.4\,\%
                & 354 & 2.7\,\% & 9913 & 19.7\,\%\\

        \hline
    \bf{Winter}      & & & & & & & & & &
                & 30020 &  19.0\,\%\\
        \hline

    April       & -- & -- & 2414 & 21.6\,\% & 7283 & 66.3\,\% & 3562 & 33.2\,\%
                & 0 & 0.0\,\% & 13259 & 30.3\,\% \\
    May         & -- & -- & 6164 & 59.1\,\% & 6846 & 65.9\,\% & 7366 & 70.9\,\%
                & 6106 & 58.7\,\% & 26482 & 63.6\,\% \\
    June        & -- & -- & 5006 & 52.3\,\% & 5516 & 57.5\,\% & 3266 & 34.2\,\%
                & 3274 & 34.2\,\% & 17062 & 44.5\,\% \\

      \hline
    \bf{Spring}      & & & & & & & & & &
                &  56803 &  46.1\,\%\\
      \hline

    July        & -- & -- & 463 & 4.6\,\% & 4509 & 44.8\,\% & 2294 & 22.8\,\%
                & 0 & 0.0\,\% & 7266 & 18.1\,\%\\
    August      & -- & -- & 0 & 0.0\,\% & 5816 & 53.3\,\% & 5837 & 53.6\,\%
                & 0 & 0.0\,\% & 11653 & 26.7\,\% \\
    September   & -- & -- & 0 & 0.0\,\% & 7641 & 65.2\,\% & 0 & 0.0\,\%
                & -- & -- & 7641 & 21.7\,\% \\

      \hline
    \bf{Summer}      & & & & & & & & & &
                & 26560  &  22.2\,\%\\
      \hline

    October     & 5159 & 38.9\,\% & 336 & 2.5\,\% & 7223 & 54.7\,\% & 0 &
                0.0\,\% & -- & -- & 12718 & 24.0\,\%\\
    November    & 3436 & 25.5\,\% & 7428 & 53.7\,\% & 6786 & 49.3\,\% & 263 & 1.9\,\%
                & -- & -- & 17913 & 32.6\,\%\\
    December    & 0 & 0.0\,\% & 3908 & 26.6\,\% & 4687 & 31.8\,\% & 0 &
                0.0\,\% & -- & -- & 8595 & 14.6\,\%\\

      \hline
    \bf{Autumn}      & & & & & & & & & &
                &  39226 &  23.7\,\%\\
      \hline

    \bf{Total}  & 8595 & 21.5\,\%  & 25960 & 18.5\,\% & 66087 & 46.9\,\% & 39697 & 28.8\,\%
                & 12270 & 14.5\,\%  &  152609 &  27.9\,\% \\
      \hline

\end{tabular}
\end{table*}


\begin{table*}
\caption{Amount of data collected and coverage percentage for
simultaneous MASS\,--\,DIMM.} \label{tab:AllDataCovSimul}
\centering
\begin{tabular}{lrrrrrrrrrrrr}
        \hline
        \multicolumn{13}{c}{\bf{Year}}           \\
        \hline
    Month   & \multicolumn{2}{c}{2004} & \multicolumn{2}{c}{2005} & \multicolumn{2}{c}{2006}
            & \multicolumn{2}{c}{2007} &  \multicolumn{2}{c}{2008}
            & \multicolumn{2}{c}{\bf{Total}}\\
    \hline

    January     & -- & --   & 63 & 0.4\,\% & 2613 & 17.9\,\% & 5451 & 37.5\,\%
                & 0 & 0.0\,\%   & 8127 & 14.0\,\%\\
    February    & -- & --   & 0 & 0.0\,\% & 2394 & 18.2\,\% & 6152 & 48.3\,\%
                & 2490 & 20.0\,\% & 11036 & 21.6\,\%\\
    March       & -- & --   & 0 & 0.0\,\% & 4521 & 36.5\,\% & 4713 & 37.0\,\%
                & 352 & 2.7\,\% & 9586 & 19.1\,\%\\

    \hline
    \bf{Winter}      & & & & & & & & & &
                &  28749 &  18.2\,\%\\
    \hline

    April       & -- & -- & 2399 & 21.5\,\% & 7129 & 64.9\,\% & 3422 & 31.9\,\%
                & 0 & 0.0\,\% & 12950 & 29.6\,\% \\
    May         & -- & -- & 5779 & 55.4\,\% & 6707 & 64.5\,\% & 7156 & 68.9\,\%
                & 5946 & 57.2\,\% & 25588 & 61.5\,\% \\
    June        & -- & -- & 4857 & 50.7\,\% & 5331 & 55.6\,\% & 3088 & 32.3\,\%
                & 3117 & 32.5\,\% & 16393 & 42.8\,\% \\

    \hline
    \bf{Spring}      & & & & & & & & & &
                &  54931 &  44.6\,\%\\
    \hline

    July        & -- & -- & 340 & 3.4\,\% & 4134 & 41.0\,\% & 1751 & 17.6\,\%
                & 0 & 0.0\,\% & 6225 & 15.5\,\%\\
    August      & -- & -- & 0 & 0.0\,\% & 5625 & 51.5\,\% & 4511 & 41.1\,\%
                & 0 & 0.0\,\% & 10136 & 23.2\,\% \\
    September   & -- & -- & 0 & 0.0\,\% & 7360 & 62.8\,\%  & 0 & 0.0\,\%
                & -- & -- & 7360 & 20.9\,\% \\

    \hline
    \bf{Summer}      & & & & & & & & & &
                &  23721 &  19.9\,\%\\
    \hline

    October     & 4033 & 30.5\,\% & 337 & 2.5\,\% & 6941 & 52.6\,\% & 0 & 0.0\,\%
                & -- & -- & 11311 & 21.4\,\%\\
    November    & 3259 & 24.2\,\% & 7274 & 52.6\,\% & 6607 & 48.0\,\% & 241 & 1.7\,\%
                & -- & -- & 17381 & 31.6\,\%\\
    December    & 0 & 0.0\,\%   & 3819 & 25.9\,\% & 4451 & 30.2\,\% & 0 & 0.0\,\%
                & -- & --   & 8270 & 14.0\,\%\\

    \hline
    \bf{Autumn}      & & & & & & & & & &
                &  36962 &  22.3\,\%\\
    \hline

    \bf{Total}  & 7292 & 18.2\,\%  & 24868 & 17.7\,\% & 63813 & 45.3\,\% & 36485 & 26.4\,\%
                & 11905 & 14.1\,\% & 144363 &  26.4\,\% \\
    \hline

\end{tabular}
\end{table*}


\subsection{Seeing Instrumentation}
\label{subsec:seeinginstr}

The SPM TMT T4 site testing station was equipped with a 35\,cm
telescope mounted approximately 7\,m above ground. At the
Cassegrain focus of the telescope a combined instrument was
deployed, which included a DIMM and a MASS, hereafter called
MASS\,--\,DIMM described by \citet{2007MNRAS.382.1268K}. Combined
MASS\,--\,DIMM instruments are extensively used for seeing
measurements and optical turbulence profiling
\citep[e.g.][]{2003MNRAS.343..891T,2005PASP..117..395T}.

The DIMM channel measures the wavefront slope differences over two
small pupils some distance apart. This yields the whole atmosphere
seeing from the telescope level to the top of the atmosphere
\citep{1990A&A...227..294S}. DIMM seeing measurements
$\varepsilon_{DIMM}$ are affected by different sources of bias
such as image threshold, defocus, exposure time, photon noise, and
high-frequency vibrations that must be monitored in real time,
thus part of the measurements are rejected. In particular, for the
system used here, the precision reported is better than
0.02\,arcsec \citep{2007ApOpt..46.6460W}.

The MASS channel detects rapid variations of light intensity in
four concentric apertures using photomultipliers and reconstructs
a turbulence profile at six different heights above the telescope,
\citep{2007MNRAS.381.1179T}. MASS is not sensitive to turbulence
near the ground as this does not produce any scintillation and it
can only measure the seeing in the free atmosphere (FA), thus
excluding the ground layer, so we have $\varepsilon_{MASS} =
\varepsilon_{FA}$. From the turbulence profile the seeing
integrated from 500\,m above the telescope to the top of the
atmosphere is computed. MASS seeing accuracy is better than
0.05\,arcsec \citep{2008ApOpt..47.2610E}.

\subsection{SPM TMT Data Coverage}
\label{subsec:datacov}

The SPM TMT T4 station acquired MASS\,--\,DIMM data from 2004
October to 2008 August. The DIMM and MASS nightly operations
commenced one hour after sunset and ceased one hour before
sunrise. DIMM and MASS measurements were triggered simultaneously,
DIMM took integrated stellar light measurements during a time
interval of 36\,s, whereas MASS sampled light during 60\,s
periods. As the DIMM channel was used to acquire the target star,
this resulted in simultaneous MASS\,--\,DIMM measurements every 70
to 90\,s \citep{2009PASP..121..527E}. Sometimes computations
failed to provide valid results probably due to cloudy conditions
or technical problems, while sometimes the telescope system was
shutdown due to bad weather conditions.

In Fig.~\ref{fig:AllDataAmount_DIMM_MASS_Simul} we present the
total amount of DIMM, MASS and simultaneous MASS\,--\,DIMM data
for each month during the acquisition period. By simultaneous
MASS\,--\,DIMM data we mean measurements within the same minute
interval.

We have also computed the statistics of the data coverage per
night and per month. To compute the monthly temporal data coverage
we have considered the duration for each night of the year due to
the diurnal cycle variation. This yields a percentage value of the
available data for each night in a given month. A 100\,\% value
would indicate data obtained during the whole night for every
night of the month.

To our knowledge, this is the first time that a complete data
coverage of seeing observations is presented in the astronomical
literature. Whereas in other papers it is spoken of number of
nights observed regardless of whether one seeing-point or many
have been observed per night, we emphasise the effective length of
time observed per night; making this a novel way of treating
seeing data.

In Tables~\ref{tab:AllDataCovDIMM}, \ref{tab:AllDataCovMASS} and
\ref{tab:AllDataCovSimul} we give the percentage of monthly,
seasonal and annual data coverage of DIMM, MASS and simultaneous
MASS\,--\,DIMM during the acquisition period.

\begin{figure}
  \begin{center}
    \includegraphics[width=\columnwidth,keepaspectratio=true]{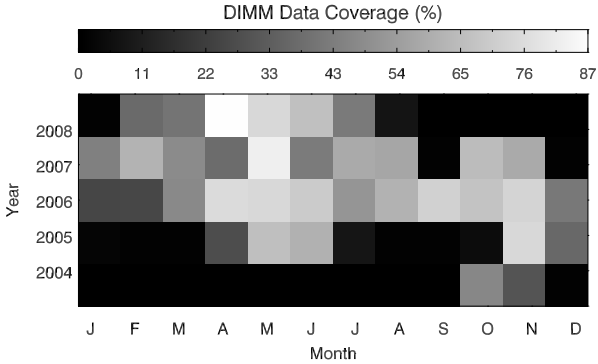}
  \end{center}
\caption{Grey level plot showing the monthly percentage of data
coverage for the 2004\,--\,2008 period for DIMM. A black square
means that no usable data were available.}
  \label{fig:DataCovDIMM}
\end{figure}

\begin{figure}
  \begin{center}
    \includegraphics[width=\columnwidth,keepaspectratio=true]{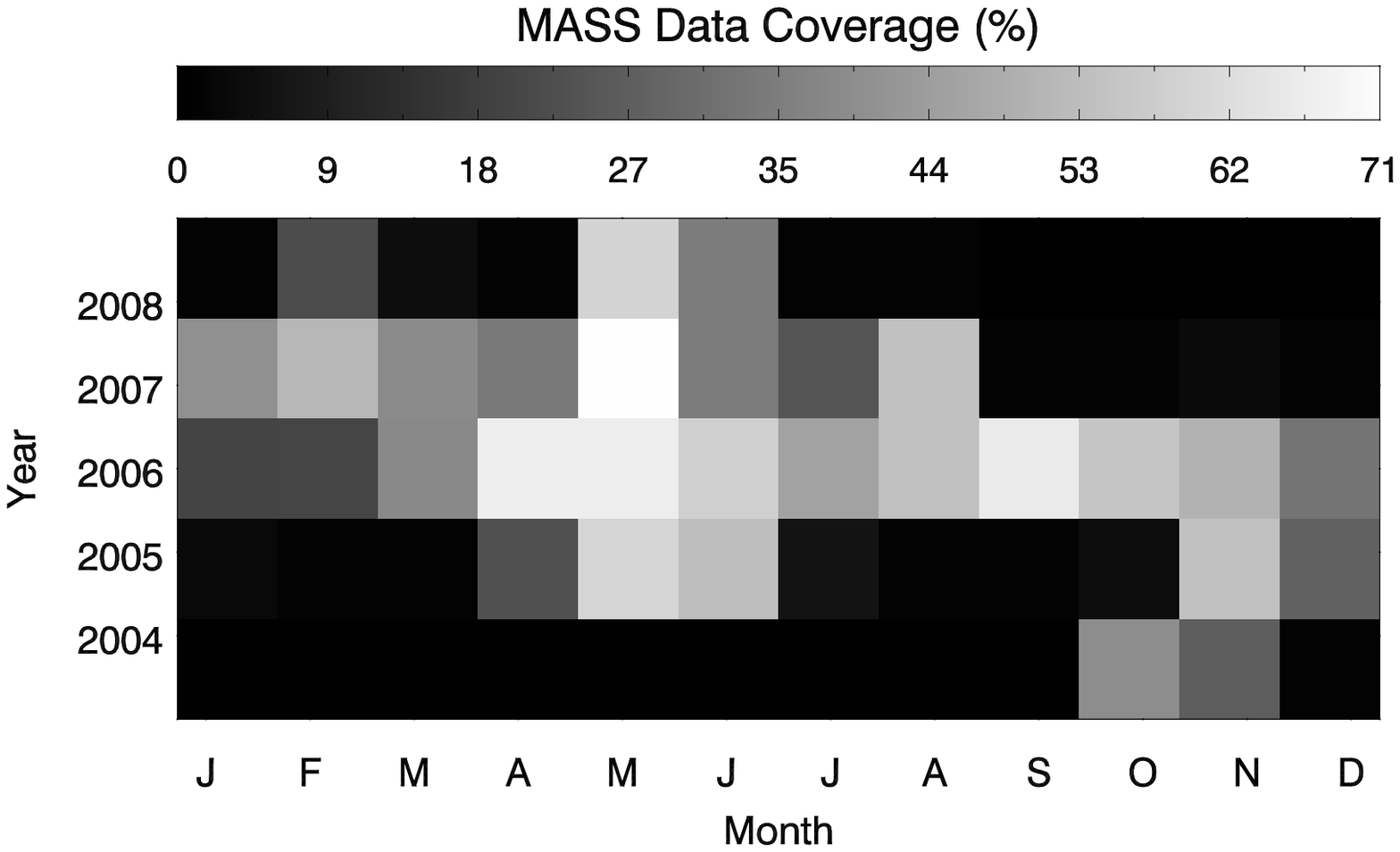}
  \end{center}
\caption{Grey level plot showing the monthly percentage of data
coverage for the 2004\,--\,2008 period for MASS. A black square
means that no usable data were available.}
  \label{fig:DataCovMASS}
\end{figure}

\begin{figure}
  \begin{center}
    \includegraphics[width=\columnwidth,keepaspectratio=true]
    {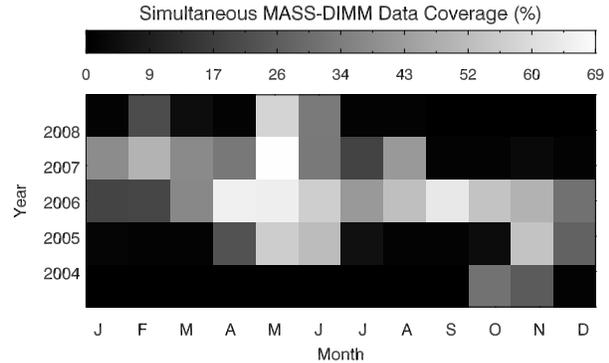}
  \end{center}
\caption{Grey level plot showing the monthly percentage of data
coverage for the 2004\,--\,2008 period for simultaneous
MASS\,--\,DIMM. A black square means that no usable data were
available.}
  \label{fig:DataCovSimul}
\end{figure}

In Figs.~\ref{fig:DataCovDIMM}, \ref{fig:DataCovMASS} and
\ref{fig:DataCovSimul} we present grey level plots showing only
the monthly percentage of data coverage for the 2004\,--\,2008
period for DIMM, MASS and simultaneous MASS\,--\,DIMM, extracted
from Tables~\ref{tab:AllDataCovDIMM}, \ref{tab:AllDataCovMASS} and
\ref{tab:AllDataCovSimul}.

These figures are intended to be a visual aid in order to see, at
a glance, which months are those with the best coverage. The
degree of darkening corresponds to the monthly percentage values
presented in Tables 2, 3 and 4.

At a first glance we immediately notice two main features. First,
that the best coverage was obtained during 2006 and 2007;
secondly, that April, May and June are the best covered months.

This is further supported by a careful analysis in
Tables~\ref{tab:AllDataCovDIMM}, \ref{tab:AllDataCovMASS} and
\ref{tab:AllDataCovSimul}. For 2006 and 2007 the DIMM data
coverage is 55.8 and 45.4\,\% respectively, while the MASS
coverage for these years is 46.9 and 28.8\,\% respectively. The
simultaneous MASS\,--\,DIMM coverage is 45.3 and 26.4\,\%
respectively. The overall data coverage for all years is
significantly less (DIMM 40.2\,\%; MASS 27.9\,\%; and simultaneous
MASS\,--\,DIMM 26.4\,\%). Although these coverage values are low
they reflect the enormous difficulties in obtaining seeing data,
even in experiments designed to ensure high data collection
efficiencies, as is the case of the TMT site testing data \citep{
2006SPIE.6267E..56R, 2010SPIE.7736E..66S}. Note that the best
sampled year for simultaneous measurements is 2006 (45.3\,\%) and
the worst sampled one is 2008 (14.1\,\%).


Seasonal MASS\,--\,DIMM coverage in winter is 18.2\,\%, spring
44.6\,\%, summer 19.9\,\%, and autumn 22.3\,\%. Seasons have been
defined at the beginning of the month: winter (January, February
and March); spring (April, May and June); summer (July, August and
September) and autumn (October, November and December).

In the monthly case we can see that the data coverage is worse in
January (14.0\,\%), December (14.0\,\%) and July (15.5\,\%), while
the best sampled months are May (61.5\,\%), June (42.8\,\%) and
November (31.6\,\%).

\section{Results}
\label{sec:results}

We have calculated the seeing from MASS, DIMM and simultaneous
MASS\,--\,DIMM obtaining nightly, monthly, seasonal, annual and
global statistics (first quartile, median and third quartile, via
cumulative distributions). The behaviour of the measured seeing
expressed as log($\varepsilon$) was verified to follow normal
statistics, obtaining an expected correlation between seeing
coverage percentage and deviations from a log-normal distribution.
The seeing distribution is expected to be log-normal because
$\varepsilon$ is a random variable that ranges from 0 to infinity,
and \citet{1974ApOpt..13.2620F} confirmed $r_{o}$ as a
log-normally distributed random variable.


\begin{figure}
  \begin{center}
    \includegraphics[width=\columnwidth]{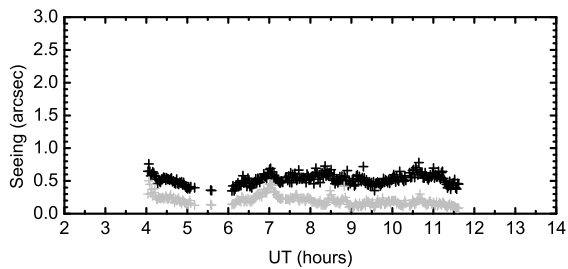}
  \end{center}
\caption{Example of DIMM (upper curve) and MASS (lower curve)
seeing median values during a good steady night (2007-08-12).}
  \label{fig:DIMM_MASS_Best_Night}
\end{figure}


\begin{figure}
  \begin{center}
    \includegraphics[width=\columnwidth]{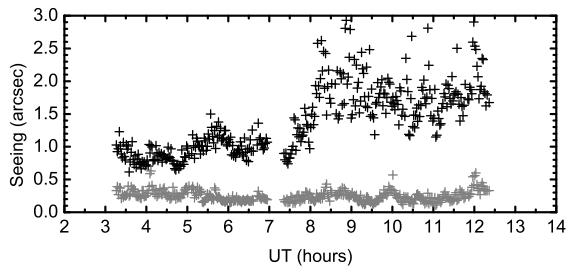}
  \end{center}
\caption{Example of DIMM (upper curve) and MASS (lower curve)
seeing median values during a night in which only DIMM seeing
degrades along the night (2006-03-21).}
  \label{fig:DIMM_MASS_Degrad_Night}
\end{figure}


\subsection{Nightly statistics}
\label{subsec:nightly}

We have made a nightly analysis of all available data ($\sim$800
nights of DIMM data, $\sim$620 nights of MASS data and $\sim$600
nights of simultaneous data, out of  $\sim$1400 nights which
corresponds to the total length of the campaign), which are either
partially covered or fully covered nights. We found a great
variety of both DIMM and MASS behaviours; nights with excellent
seeing throughout the night; erratic seeing nights; degrading
seeing in which there is a sudden burst of bad seeing; etc. In the
following paragraphs we will discuss a couple of examples.


\begin{figure*}
  \begin{center}
    \includegraphics[width=16cm]{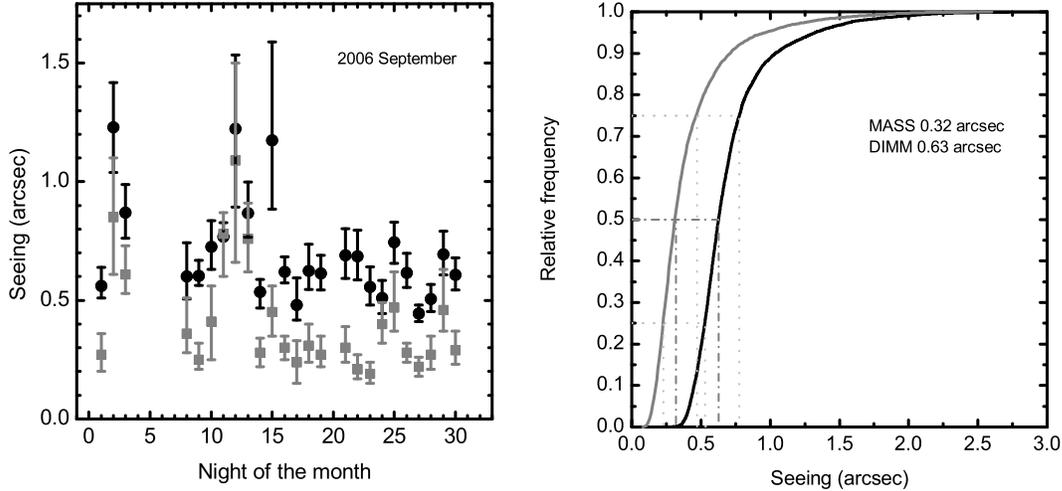}
  \end{center}
\caption{Left panel: DIMM (circles) and MASS (squares) statistics
for simultaneously obtained data. Median seeing values (with lower
and upper limits represented by first and third quartiles
respectively) for each night of the month 2006 September. Right
panel: Seeing Cumulative Distribution Function for the same month.
Left curve represents MASS values with a median of 0.32\,arcsec.
Right curve represents DIMM values with a median of 0.63\,arcsec.}
  \label{fig:DIMM_MASS_Best_Month1}
\end{figure*}


In Fig.~\ref{fig:DIMM_MASS_Best_Night} we present a night (2007
August 12) with good steady seeing. For both DIMM and MASS the
seeing variations are well correlated. The median seeing measured
by DIMM and MASS is 0.53 and 0.19\,arcsec, respectively.

In Fig.~\ref{fig:DIMM_MASS_Degrad_Night} we present a night (2006
March 21) in which it is clear that there is a strong lack of
correlation between the DIMM and MASS seeing. Only the DIMM seeing
degraded during the night, while the MASS seeing remains steady
with a median of 0.24\,arcsec.


\begin{figure*}
  \begin{center}
    \includegraphics[width=16cm]{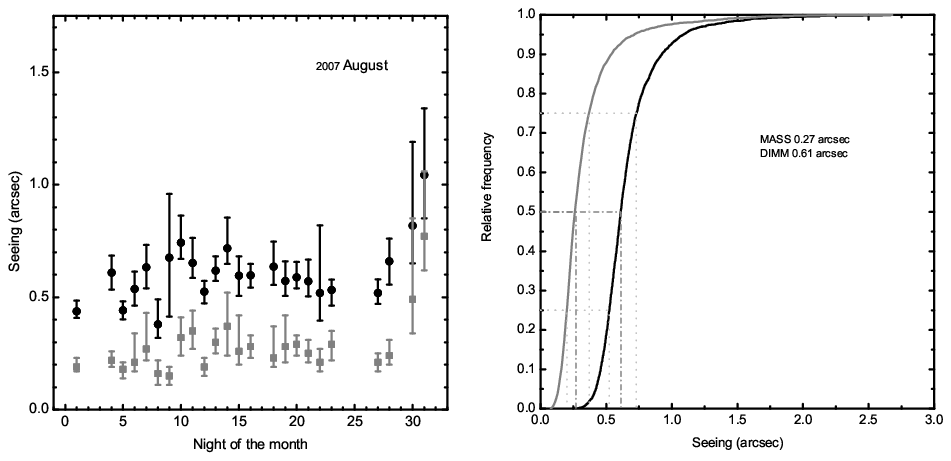}
  \end{center}
\caption{ Left panel: DIMM (circles) and MASS (squares) statistics
for simultaneously obtained data. Median seeing  values (with
lower and upper limits represented by first and third quartiles
respectively) for each night of the month 2007 August. Right
panel: Seeing Cumulative Distribution Function for the same month.
Left curve represents MASS values with a median of 0.27\,arcsec.
Right curve represents DIMM values with a median of 0.61\,arcsec.}
  \label{fig:DIMM_MASS_Best_Month2}
\end{figure*}


The median standard variation of the whole atmosphere seeing
within a night is 0.19\,arcsec (with first and third quartiles
0.12 and 0.29\,arcsec, respectively).

As an example of simultaneous MASS\,--\,DIMM nightly statistics we
show in the left panel of Figs.~\ref{fig:DIMM_MASS_Best_Month1}
and \ref{fig:DIMM_MASS_Best_Month2}, the seeing median values for
some of the best sampled months (2006 September with 25 nights,
and 2007 August with 24 nights).


\subsection{Monthly statistics}
\label{subsec:monthly}


\begin{table*}
\caption{Median seeing monthly values (arcsec).}
\label{tab:MonthlySeeingDIMMMASS} \centering
\begin{tabular}{lrrrrrrrrrrrr}
        \hline
        \multicolumn{13}{c}{\bf{Year}}           \\
        \hline
    Month   & \multicolumn{2}{c}{2004} & \multicolumn{2}{c}{2005} & \multicolumn{2}{c}{2006}
            & \multicolumn{2}{c}{2007} &  \multicolumn{2}{c}{2008}
            & \multicolumn{2}{c}{\bf{Total}}\\
    \hline
            & \multicolumn{1}{c}{DIMM} & \multicolumn{1}{c}{MASS} & \multicolumn{1}{c}{DIMM}
            & \multicolumn{1}{c}{MASS} &  \multicolumn{1}{c}{DIMM}
            & \multicolumn{1}{c}{MASS} & \multicolumn{1}{c}{DIMM} & \multicolumn{1}{c}{MASS}
            & \multicolumn{1}{c}{DIMM} &  \multicolumn{1}{c}{MASS}
            & \multicolumn{1}{c}{DIMM} & \multicolumn{1}{c}{MASS}\\
        \hline
    January     & -- & --   & 1.47 & 0.82 & 0.75 & 0.36 & 1.10 & 0.47
                & -- & --   & 1.01 & 0.44\\
    February    & -- & --   & --  & -- & 1.07 & 0.52 & 0.98 & 0.41
                & 0.86 & 0.34 & 0.99 & 0.41\\
    March       & -- & --   & -- & -- & 1.08 & 0.33 & 0.82 & 0.39
                & 0.96 & 0.50 & 0.92 & 0.37\\
    April       & -- & -- & 0.98 & 0.40 & 0.83 & 0.37 & 0.92 & 0.45
                & -- & -- & 0.88 & 0.40\\
    May         & -- & -- & 0.77 & 0.28 & 0.64 & 0.30 & 0.76 & 0.30
                & 0.89 & 0.43 & 0.75 & 0.32\\
    June        & -- & -- & 0.84 & 0.30 & 0.74 & 0.41 & 0.73 & 0.39
                & 0.65 & 0.33 & 0.75 & 0.36\\
    July        & -- & -- & 0.94 & 0.59 & 0.72 & 0.46 & 0.58 & 0.30
                & -- & -- & 0.68 & 0.41\\
    August      & -- & -- & -- & -- & 0.64 & 0.35 & 0.61 & 0.27
                & -- & -- & 0.63 & 0.31\\
    September   & -- & -- & -- & -- & 0.62 & 0.31 & -- & --
                & -- & -- & 0.62 & 0.31\\
    October     & 1.08 & 0.46 & 0.85 & 0.33 & 0.75 & 0.38 & -- & --
                & -- & -- & 0.83 & 0.40\\
    November    & 0.97 & 0.41 & 0.83 & 0.42 & 0.83 & 0.40 & 0.89 & 0.52
                & -- & -- & 0.85 & 0.41\\
    December    & -- & --   & 0.78 & 0.38 & 0.82 & 0.38 & -- & --
                & -- & --   & 0.80 & 0.38\\
        \hline

\end{tabular}
\end{table*}


\begin{figure*}
  \begin{center}
    \includegraphics[width=16cm]{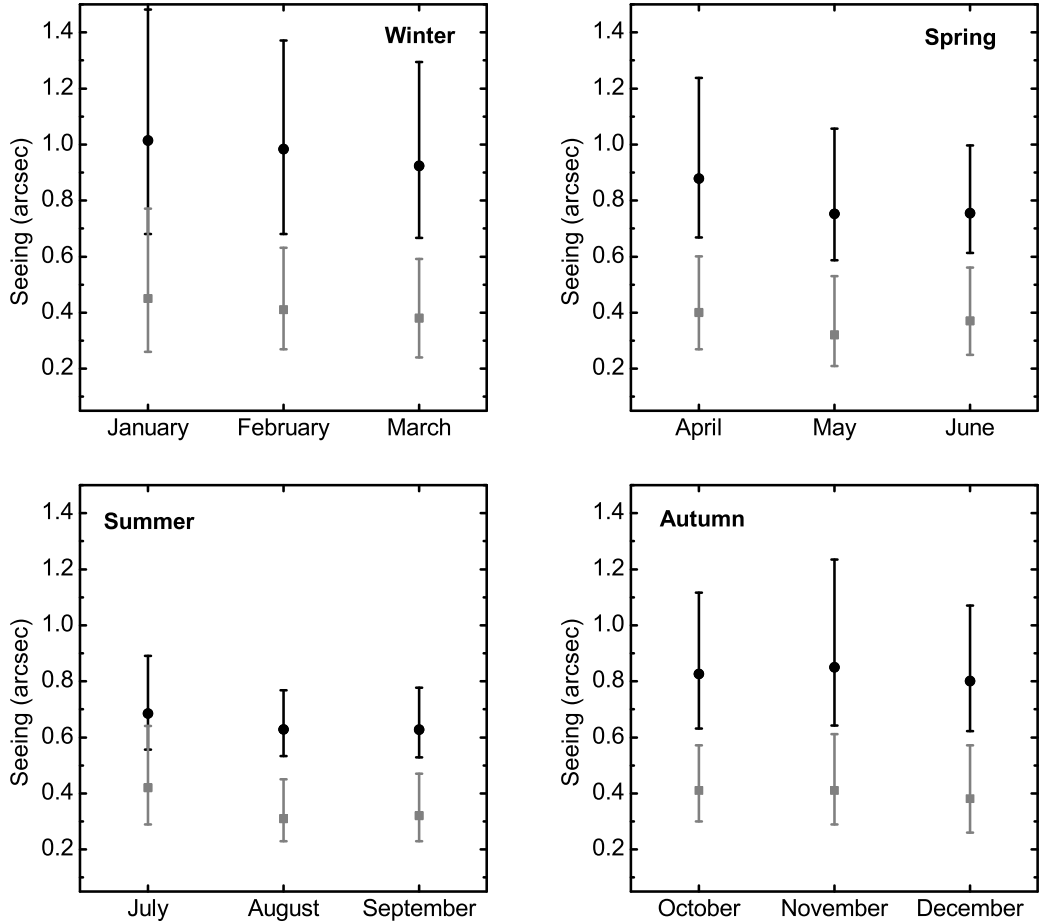}
  \end{center}
\caption{Median seeing values (with first and third quartiles as
error bars) for each month of every season covered by the campaign
(2004\,--\,2008). Symbols as in
Fig.~\ref{fig:DIMM_MASS_Best_Month1} left panel.}
  \label{fig:4Seasons}
\end{figure*}


\begin{table*}
\caption{Seasonal and annual seeing statistics (arcsec).}
\label{tab:DIMM_MASS_AllSeasonsYears} \centering
\begin{tabular}{lcccccccccr}

\hline
        & \multicolumn{3}{c}{DIMM} & \multicolumn{3}{c}{MASS} & \multicolumn{3}{c}{Ground Layer} & \\
\hline
   Season &  1st &  Median & 3rd   & 1st  & Median & 3rd & 1st & Median & 3rd & Coverage\\
          & quartile &  & quartile & quartile & & quartile & quartile & & quartile & \% \\
\hline
    Winter & 0.67 & 0.97 & 1.38 & 0.26 & 0.40 & 0.64 & 0.51 & 0.73 & 1.11 & 18.2\\
    Spring & 0.61 & 0.78 & 1.07 & 0.23 & 0.35 & 0.54 & 0.46 & 0.59 & 0.81 & 44.6\\
    Summer & 0.54 & 0.64 & 0.79 & 0.24 & 0.33 & 0.48 & 0.38 & 0.46 & 0.56 & 19.9\\
    Autumn & 0.64 & 0.83 & 1.16 & 0.28 & 0.40 & 0.58 & 0.46 & 0.63 & 0.92 & 22.3\\
\hline
Year  & & &   & & &  & & \\
\hline
    2004 & 0.72 & 1.04 & 1.51 &  0.31 & 0.44 & 0.65 & 0.56 & 0.81 & 1.27 & 18.2\\
    2005 & 0.65 & 0.82 & 1.13 &  0.24 & 0.36 & 0.54 & 0.49 & 0.65 & 0.89 & 17.7\\
    2006 & 0.58 & 0.74 & 1.02 &  0.25 & 0.37 & 0.55 & 0.42 & 0.54 & 0.76 & 45.3\\
    2007 & 0.60 & 0.80 & 1.17 &  0.24 & 0.37 & 0.58 & 0.46 & 0.60 & 0.87 & 26.4\\
    2008 & 0.61 & 0.82 & 1.11 &  0.26 & 0.38 & 0.57 & 0.45 & 0.62 & 0.86 & 14.1\\
\hline

\end{tabular}
\end{table*}


In Table~\ref{tab:MonthlySeeingDIMMMASS} we present detailed
monthly statistics for DIMM and MASS data for the whole campaign.

Apart from the worst median seeing value obtained in 2005 January,
which happens to be the month with the poorest coverage (0.4\,\%,
c.f. Table~\ref{tab:AllDataCovSimul}), DIMM seeing values vary
from 0.6 to 1.1\,arcsec and MASS seeing values vary from 0.3 to
0.4\,arcsec. An example of monthly statistics is shown in the
right panels of Figs.~\ref{fig:DIMM_MASS_Best_Month1} and
\ref{fig:DIMM_MASS_Best_Month2} containing the seeing Cumulative
Distribution Functions (CDF) for the same months previously
mentioned. For DIMM we obtain a median seeing of 0.62 and
0.61\,arcsec, respectively; while for MASS we get 0.31 and
0.27\,arcsec, respectively.


\subsection{Seasonal statistics}
\label{subsec:seasonalstat}

We show in Fig.~\ref{fig:4Seasons} the seeing seasonal behaviour.
For each season we plot the DIMM and MASS median seeing values for
the corresponding months. The DIMM seeing displays a pronounced
seasonal variability and there seems to be a slight correlation
between the DIMM and MASS variations. It is clear that the worst
whole atmosphere seeing occurs in winter, although at the end of
the season there is a tendency towards a better seeing. This
tendency continues until the middle part of summer when the best
seeing is achieved, worsening progressively towards autumn.

In the upper part of Table~\ref{tab:DIMM_MASS_AllSeasonsYears} we
present the seasonal statistics for the entire observing run. As
mentioned above the worst whole atmosphere seeing occurs in winter
(0.97\,arcsec) and the best in summer (0.64\,arcsec).

Differences in seasonal MASS are smaller indicating that most of
the large DIMM seeing variations are due to the turbulence
occurring near the ground (see Sect.~\ref{subsec:groundstat}).


\begin{figure}
  \begin{center}
    \includegraphics[width=\columnwidth]{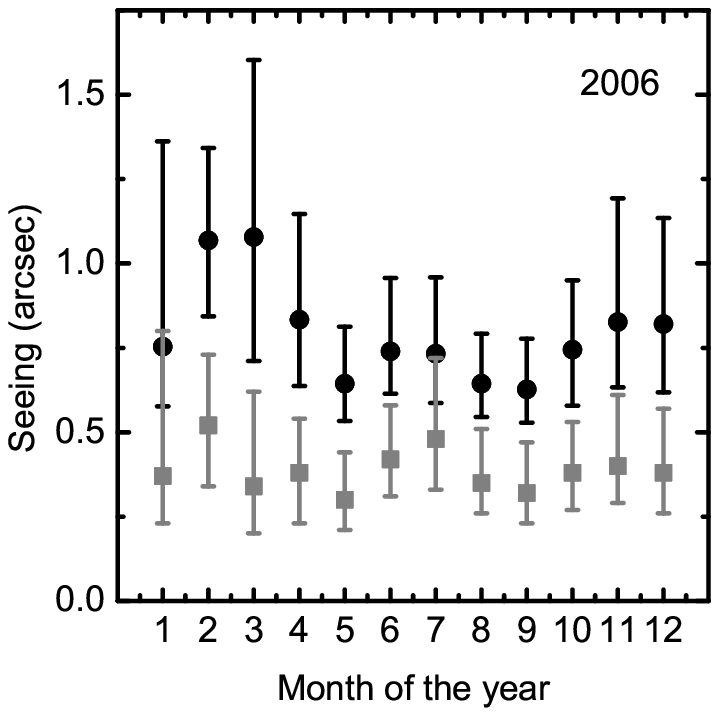}
  \end{center}
\caption{DIMM (circles) and MASS (squares) simultaneously obtained
data. Median seeing values (with first and third quartiles as
error bars) for each month of 2006.}
  \label{fig:DIMM_MASS_anual_2006}
\end{figure}


\begin{figure}
  \begin{center}
    \includegraphics[width=\columnwidth]{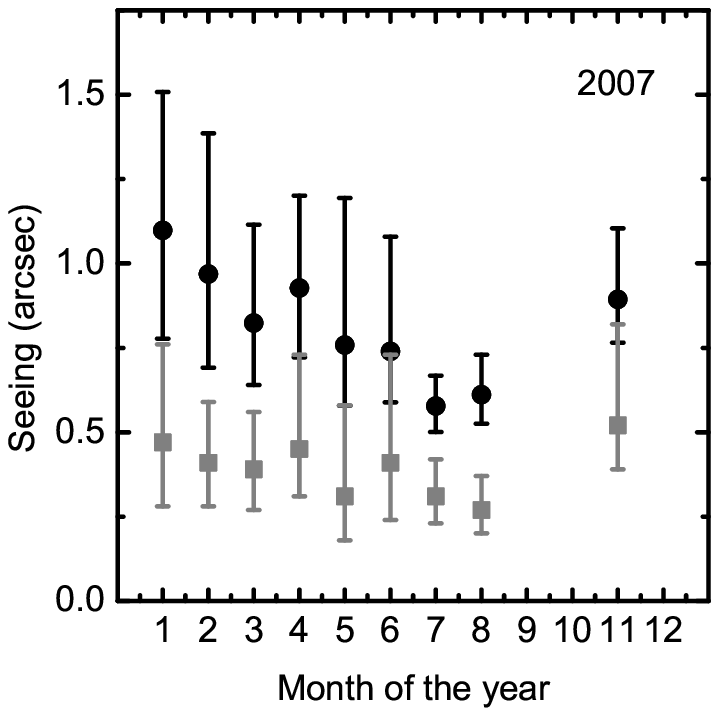}
  \end{center}
\caption{DIMM (circles) and MASS (squares) simultaneously obtained
data. Median seeing values (with first and third quartiles as
error bars) for each month of 2007.}
  \label{fig:DIMM_MASS_anual_2007}
\end{figure}


\subsection{Annual statistics}
\label{subsec:annualstat}

We present in Figs.~\ref{fig:DIMM_MASS_anual_2006} and
\ref{fig:DIMM_MASS_anual_2007} the DIMM and MASS seeing
measurements for the years 2006 and 2007 which are the best
sampled years in terms of seasonal coverage.


\begin{table}
\caption{Ground layer monthly median seeing (arcsec).}
\label{tab:GroundLayerMonthlyStatistics} \centering
\begin{tabular}{lcccccr}
        \hline
        \multicolumn{7}{c}{\bf{Year}}           \\
        \hline
        Month   & 2004   &  2005 &  2006  &  2007 &  2008 & {\bf{Total}}\\
        \hline

    January     &   --        &   0.97      &   0.60      &    0.82    &   --      &  0.76  \\
    February    &   --        &   --        &   0.77      &    0.76    &   0.72    &  0.75  \\
    March       &   --        &   --        &   0.89      &    0.61    &   0.67    &  0.69  \\
    April       &   --        &   0.78      &   0.64      &    0.65    &   --      &  0.66  \\
    May         &   --        &   0.65      &   0.50      &    0.58    &   0.65    &  0.58  \\
    June        &   --        &   0.68      &   0.53      &    0.53    &   0.49    &  0.57  \\
    July        &   --        &   0.64      &   0.45      &    0.42    &   --      &  0.45  \\
    August      &   --        &   --        &   0.46      &    0.48    &   --      &  0.47  \\
    September   &   --        &   --        &   0.47      &    --      &   --      &  0.47  \\
    October     &   0.81      &   0.72      &   0.56      &    --      &   --      &  0.63  \\
    November    &   0.80      &   0.61      &   0.62      &    0.61    &   --      &  0.64  \\
    December    &   --        &   0.58      &   0.62      &    --      &   --      &  0.60  \\

        \hline

\end{tabular}
\end{table}


In both figures we can see a clear seasonal variation of the DIMM
seeing. Whereas the MASS values tend to remain more stable
throughout the year.

The lower part of Table~\ref{tab:DIMM_MASS_AllSeasonsYears}
presents the yearly DIMM and MASS statistics. It is interesting to
note that the best whole atmosphere seeing occurs for 2006
(0.74\,arcsec) which is the best sampled year.


\subsection{Ground layer statistics}
\label{subsec:groundstat}

We calculated the Ground layer (GL) seeing $\varepsilon_{GL}$
(from 7\,--\,500\,m) which is defined as the difference between
the DIMM and MASS seeing. As we note from
equation~\ref{eq:seeingeq} the contributions of the different
layers to the seeing must be summed as the 5/3 power, which yields
\begin{equation}
   \varepsilon_{GL} = (\varepsilon^{5/3}_{DIMM} - \varepsilon^{5/3}_{MASS})^{3/5}.
\label{eq:groundlayereq}
\end{equation}

{\bf In the data acquisition process it might occur that
$\varepsilon_{DIMM}\,<\,\varepsilon_{MASS}$. These occurrences
correspond to what \citet{2007RMxAC..31...61T} defined as
``over-shoots''. We identified that the total number of these
occurrences corresponds to $\sim1.5\,\%$. We have decided to
reject these data noting that the amount of rejected data is
rather small and therefore overall statistics of the GL seeing are
not affected.}

In the GL section of Table~\ref{tab:DIMM_MASS_AllSeasonsYears} we
present the seasonal and yearly results for the GL seeing values.
While in Table~\ref{tab:GroundLayerMonthlyStatistics} we show the
monthly variation of the GL seeing for the entire observing run.
We note that the best GL seeing is during July, August and
September months with a value $\sim$0.46\,arcsec.

Again, the GL behaviour is similar to the DIMM seeing: best in
summer, worst in winter. This is due to the fact that the MASS
values are always nearly constant showing that the high-altitude
turbulence at SPM is small. Therefore, the main contribution to
the whole atmosphere seeing at SPM comes from a strong GL.


\subsection{Global statistics and hourly trend}
\label{subsec:globalstat}

\begin{figure*}
    \begin{center}
        \includegraphics[width=16cm]{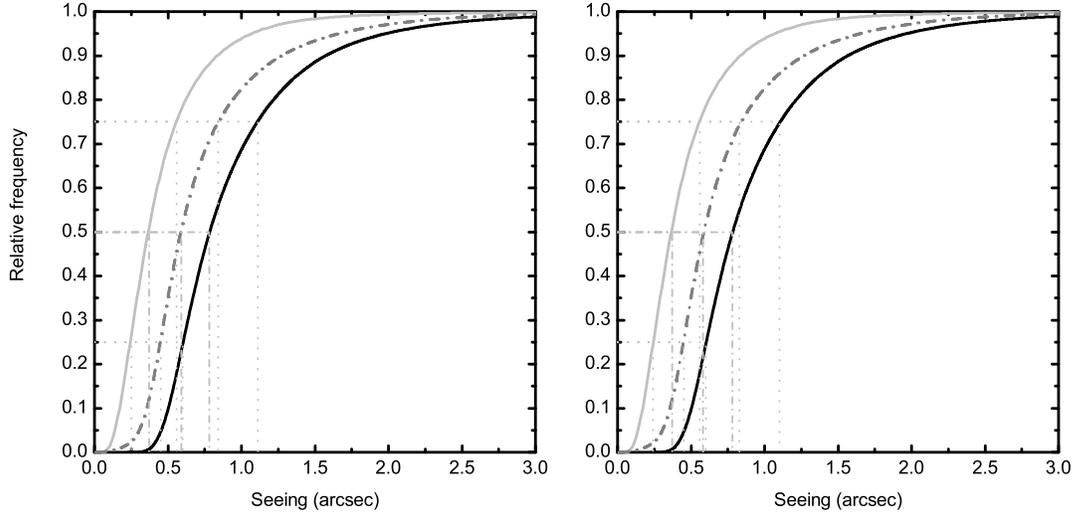}
    \end{center}
\caption{MASS, DIMM and GL statistics for simultaneously obtained
data. Left panel: Cumulative Distribution Functions (CDFs) for the
whole campaign (2004\,--\,2008) including all data points.
Median(MASS)\,=\,0.37\,arcsec, Median(DIMM)\,=\,0.78\,arcsec,
Median(GL)\,=\,0.59\,arcsec. Right panel: CDFs for the whole
campaign (2004\,--\,2008) including only 4--11 UT data points.
Median(MASS)\,=\,0.37\,arcsec, Median(DIMM)\,=\,0.77\,arcsec,
Median(GL)\,=\,0.58\,arcsec. Left curves represents MASS values,
middle curves represents GL values and right curves represents
DIMM values.}
\label{fig:CDF_Global_Simult_Data_Hours_4_11_DIMM_MASS_GL}
\end{figure*}

In the left panel of
Fig.~\ref{fig:CDF_Global_Simult_Data_Hours_4_11_DIMM_MASS_GL} we
show the CDF for the DIMM, MASS, and derived GL seeing for all
simultaneous data. Table~\ref{tab:GlobalStats} shows the overall
seeing results.

\begin{figure}
\begin{center}
    \includegraphics[width=\columnwidth]{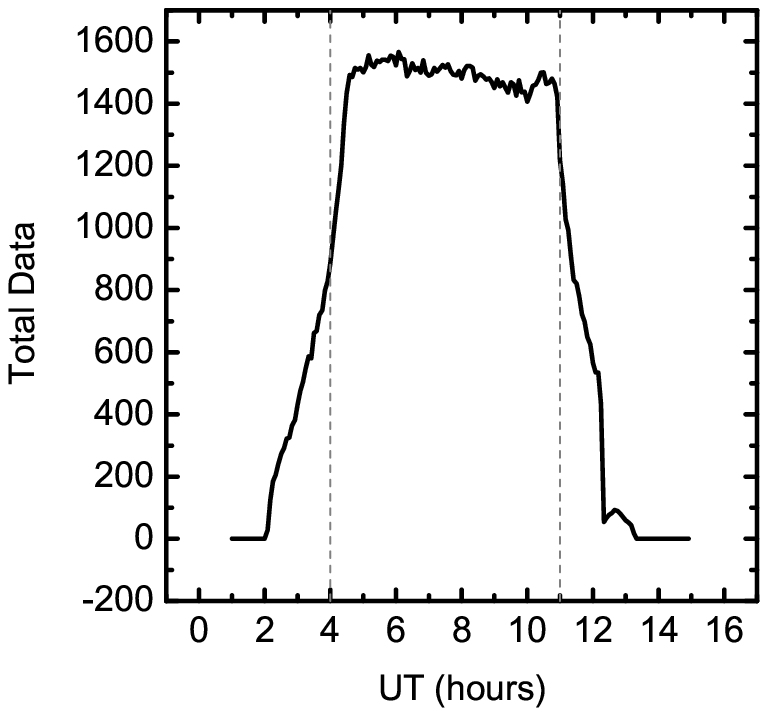}
  \end{center}
\caption{Total number of hourly simultaneous DIMM and MASS
measurements (N). The vertical lines delimit the interval between
4 and 11 UT hours.}
  \label{fig:totalhourlydata}
\end{figure}

\citet{2009PASP..121.1151S} argue that the DIMM seeing at SPM
rises during the second half of the night (see their Fig.~4). We
decided to explore the seeing behaviour and look for a possible
degradation along the night. In Fig.~\ref{fig:totalhourlydata} we
have plotted the number of data as a function of the hour at which
they were taken. It is clear that before 4 UT hours and after 11
UT hours the number of measurements taken was substantially
smaller than within 4 to 11 UT. This justifies the calculation of
an overall CDF in this restricted range presented on the right
panel of
Fig.~\ref{fig:CDF_Global_Simult_Data_Hours_4_11_DIMM_MASS_GL}. The
second line of Table~\ref{tab:GlobalStats} shows the overall
results only for the same interval. As it is clear, the resulting
statistics does not significantly change when using the restricted
time interval.

\begin{figure}
\begin{center}
    \includegraphics[width=\columnwidth]{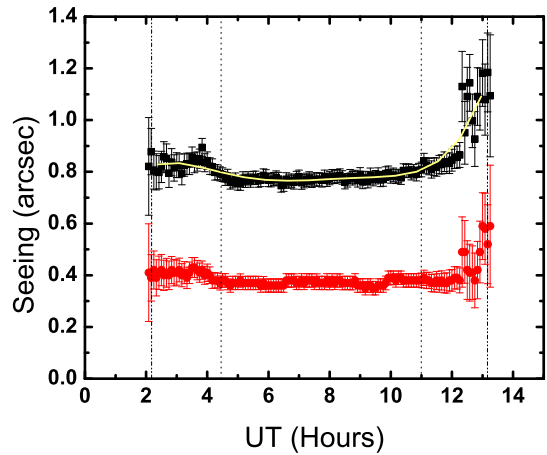}
  \end{center}
\caption{Hourly seeing computed from total simultaneous DIMM
(upper curve) and MASS (lower curve) data. The error bars
correspond to $\frac{1}{\sqrt{N}}$. {\bf The fitted curve to the
DIMM data corresponds to a weighted fifth degree polynomial.
Vertical dot-dot lines indicate the shortest night, whereas
vertical dash-dot-dot lines delimit the longest night of the
year.}}
  \label{fig:hourlyseeing}
\end{figure}

In Fig.~\ref{fig:hourlyseeing} we plot the DIMM and MASS hourly
results integrated over the whole campaign. {\bf The seeing seems to
be worse at the beginning and at the end of the night, however the number
of data (N) which supports this assertion is rather small and it is
clear from the figure that these data correspond to the longest nights of the
year (i.e. winter nights where seeing is usually worse). The
points on the plot have error bars corresponding to
$\frac{1}{\sqrt{N}}$.

We have made a weighted fifth grade polynomial fit to the
points. It is clear that in the interval delimited by the dot-dot
lines (corresponding to the shortest night of the year) the fit is
essentially linear and horizontal, which might indicate that the
seeing remains constant throughout the night.} We know that in
giving the data points an error bar equal to $\frac{1}{\sqrt{N}}$
those points within the 4 to 11 UT interval, where N is
sufficiently large, have a determinant influence on the results of
the fit. However, it is correct to do it this way since we have
shown that the statistical results in the 4 to 11 UT interval are
the same as those for the whole interval (see
Table~\ref{tab:GlobalStats}).


\begin{figure}
\begin{center}
    \includegraphics[width=\columnwidth]{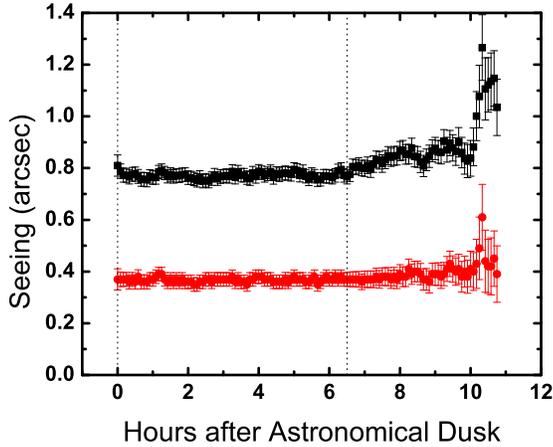}
  \end{center}
\caption{{\bf Hourly seeing computed from total simultaneous DIMM
(upper curve) and MASS (lower curve) data, for hours after
astronomical dusk. The error bars correspond to
$\frac{1}{\sqrt{N}}$. Vertical dot-dot lines delimit the shortest
night of the year.}}
  \label{fig:hourlyseeing_dusk}
\end{figure}

\begin{figure}
\begin{center}
    \includegraphics[width=\columnwidth]{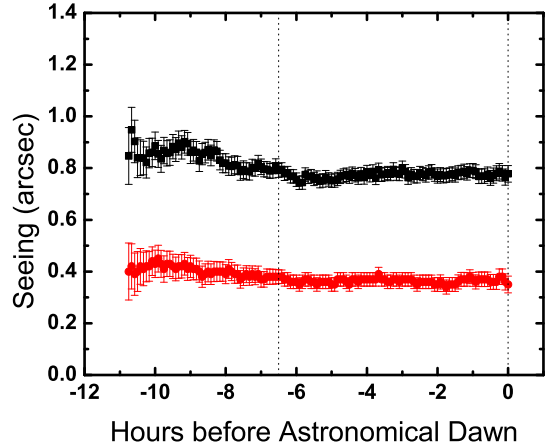}
  \end{center}
\caption{{\bf Hourly seeing computed from total simultaneous DIMM
(upper curve) and MASS (lower curve) data, for hours before
astronomical dawn. The error bars correspond to
$\frac{1}{\sqrt{N}}$. Vertical dot-dot lines delimit the shortest
night of the year.}}
  \label{fig:hourlyseeing_dawn}
\end{figure}

{\bf In order to explore with precision the seeing behaviour at
the beginning and the end of the night we analyse the data in a
different manner: integrating with respect to the astronomical
dusk and to the astronomical dawn, calculated for each night of
the year. We do this since the beginning and the end of the night
occur at a different time (UT) throughout the year.

We include Figs.~\ref{fig:hourlyseeing_dusk} and
\ref{fig:hourlyseeing_dawn} which show the median seeing versus
hours after astronomical dusk and before astronomical dawn.

Performing a linear fit to the DIMM points contained within the
length of the shortest night we find it to be essentially flat
with slopes ${9.0311\times10^{-4}}\pm{0.0016}$ for
Fig.~\ref{fig:hourlyseeing_dusk} and
${0.0024}\pm{0.0016}$ for
Fig.~\ref{fig:hourlyseeing_dawn}.

The MASS points do not seem to show any significant tendency in
either case with slopes of ${3.9400\times10^{-4}}\pm0.0016$ and
${-2.4118\times10^{-4}}\pm0.0016$.

We, therefore, conclude that the seeing at SPM does not
statistically change either shortly after dusk or towards the last
few hours of the night, in contrast with the results reported by
\citet{2009PASP..121.1151S}, and in good agreement with the
findings of \citet{1998RMxAA..34...47E} and
\citet{2003RMxAC..19...23S} where no clear tendency was found.}


\begin{table*}
\caption{Global seeing statistics (arcsec).}
\label{tab:GlobalStats} \centering
\begin{tabular}{lccccccccc}

\hline
        & \multicolumn{3}{c}{DIMM} & \multicolumn{3}{c}{MASS} & \multicolumn{3}{c}{Ground Layer} \\
\hline
   Data &  1st &  Median & 3rd   & 1st  & Median & 3rd & 1st & Median & 3rd \\
          & quartile &  & quartile & quartile & & quartile & quartile & & quartile \\
\hline

    All         & 0.60 & 0.78 & 1.11 & 0.25 & 0.37 & 0.56 & 0.45 & 0.59 & 0.84 \\
    4--11 UT    & 0.60 & 0.78 & 1.10 & 0.24 & 0.37 & 0.56 & 0.45 & 0.58 & 0.83 \\

\hline

\end{tabular}
\end{table*}


\subsection{Comparison with previous results}
\label{subsec:compprev}

Our seasonal results (see Sect.~\ref{subsec:seasonalstat}) may be
compared with the findings of \citet{1998RMxAA..34...47E} with the
STT and CM monitors and by \citet{2003RMxAA..39..291M} with a DIMM
monitor. The former found a summer median of 0.58 arcsec while the
latter found a median of 0.55 arcsec, compared with 0.64\,arcsec
obtained here. Although, we find higher values in all seasons in
this paper, summer is still shown to be the best season in all
works. We point out that the coverage percentage in summer, autumn
and winter, presented in this paper are particularly low ($\leq
23\%$). Still the median value for spring is much higher than
previous works; 0.60\,arcsec in \citet{1998RMxAA..34...47E} and
0.61\,arcsec in \citet{2003RMxAA..39..291M} and 0.78 (this paper).
The results for autumn are 0.68, 0.63 and 0.83\,arcsec
respectively, while for winter are 0.69, 0.77 and 0.97\,arcsec
respectively. Thus, the results are consistent, if not in the
absolute values, in the fact that summer yields the best values,
while winter is the worst season. The mean values for all three
campaigns yield 0.79, 0.68, 0.61 and 0.70\,arcsec for winter,
spring, summer and autumn respectively, while the overall median
average is 0.70\,arcsec. This value is higher than the median
seeing in the first two campaigns and lower than the results in
this work (see Sect.~\ref{subsec:globalstat}). The differences
show the skewness produced by uneven coverage during the year in
all three campaigns.

Our global seeing results are in good agreement with those of
\citet{2009PASP..121.1151S}. The slight differences found could be
due to a larger temporal coverage in our data set, from 2004 Oct.
to 2008 Aug.

Comparing the results in this work (Table~\ref{tab:GlobalStats})
with those of previous studies (Table~\ref{tab:ResultsSeeing} and
Sect.~\ref{subsec:seasonalstat}) reveals that the whole atmosphere
seeing ($\varepsilon_{DIMM}$) is larger than most of the values
reported before. This could be due to a slight degradation in
seeing in recent years.

Based on the results for Paranal and La Silla
\citep{2008Msngr.132...11S}\footnote{Sarazin
M.,~2010,~http://www.eso.org/gen-fac/pubs/astclim/paranal/seeing/singstory.html}
we suggest that the apparent seeing degradation found at SPM might
be correlated with climate change. We have consulted the Climatic
Data Base of the Northwest Mexico
(\url{http://peac-bc.cicese.mx/datosclim}) which contains weather
data for a number of stations near the SPM observatory site for
the years 1981 to 2008. For this time period, we found that the
temperature tends to increase 0.05 Celsius per year while the
water precipitation tends to decrease 5 mm per year. These
variations clearly point to a climate change which might produce
local effects. Whether the seeing degradation is due to these
effects is yet to be established. However, a full study of the
dependence of the seeing value with the SPM climate change is
beyond the scope of this paper.

From a statistical point of view we found that the seeing at SPM
does not suffer any degradation towards dawn, in contrast with the
results reported by \citet{2009PASP..121.1151S}. However, the
consistent increase of seeing values towards the end of the night
might suggest, as \citet{2009PASP..121.1151S} states, that seeing
degrades towards dawn. To establish this it would be necessary to
collect a large number of observations towards dawn so that this
result may be established with a sufficient degree of statistical
significance.

\citet{2012MNRAS.420.1273C} made a study of SPM solar radiation
(cloud coverage) using the TMT data for the same period presented
in this paper. They state that SPM skies are clear in spring,
relatively cloudy in summer and fairly clear in winter. It is
interesting to notice that there appears to be a correlation
between the cloud cover during day time and the nightly value of
the seeing, in the sense that cloudy days have better seeing at
night and viceversa. This might indicate that the processes that
occur when the cloud cover is dissipated contribute to smaller
amounts of atmospheric turbulence. This is an interesting point
which merits further study that is beyond the scope of this work.


\section{Summary and Conclusions}
\label{sec:conclusions}

In this work we have used the data obtained by the TMT Site
Testing Project for the SPM site in the north-west of Me\-xi\-co.
The data consist in DIMM and MASS measurements taken over a period
of nearly five years (2004\,October\,--\,2008\,August, $\sim$1400
nights). A variety of previous studies have measured the seeing at
SPM. The obtained median seeing varies from 0.50 to 0.79\,arcsec;
although these measurements have been taken in different epochs
and with disparate instruments. The coverage of each observing run
is also limited.


For the analysis presented in this paper, we have made a detailed
coverage study obtaining percentages of total observed time per
night with respect to total night length. This is the first time
that this approach is presented in the seeing astronomical
literature. The total coverage for DIMM, MASS and simultaneous
data is 40.2\,\%, 27.9\,\%, and 26.4\,\% respectively. This means
that, of the 1400 available nights, the DIMM data were obtained in
an equivalent of 563 ``fully observed'' nights (nights with a
100\,\% time coverage), whereas for the MASS this corresponds to
391 nights, while for the simultaneous data it corresponds to 370
nights.

We calculated nightly, monthly, seasonal, annual and global
statistics of the seeing from MASS, DIMM and simultaneous
MASS\,--\,DIMM observations. The simultaneous MASS\,--\,DIMM
results indicate that the best seeing is obtained for 2006
September: 0.62 (DIMM),  0.31 (MASS)\,arcsec and 2007 August: 0.61
(DIMM), 0.27 (MASS)\,arcsec. These months correspond to our
definition of the summer season in which the calculated seeing is
0.64, 0.33 and 0.46\,arcsec, median values for DIMM, MASS and
ground-layer respectively. Making the summer the best seeing
season at SPM. It is worth noting that the seeing obtained for
2006 -- 0.74 (DIMM), 0.37 (MASS) and 0.54\,arcsec (GL) -- is the
best yearly seeing of the whole campaign which also corresponds to
the best sampled year. For the GL seeing the best month is July
with a value of 0.45\,arcsec. The overall results yield a median
seeing of 0.78 (DIMM), 0.37 (MASS) and 0.59\,arcsec (GL).
Comparing with previous works we found that our whole atmosphere
seeing values are slightly larger than most of the values reported
before, perhaps this is due to a degradation of seeing with time,
which might be caused by climate effects. {\bf The hourly analysis
clearly showed that there is no statistically significant tendency
of seeing degradation towards dawn, in contrast with that reported
by Skidmore (2009).}


These results show that SPM is a competitive astronomical site, in
which it would be worth performing a continuous astroclimate
evaluation.

\section*{Acknowledgments}

This paper is based on data kindly available to the public at
\url{http://sitedata.tmt.org/} and obtained by the TMT Site
Testing Project. We acknowledge the TMT team and the technical and
administrative staff of the OAN\,--\,SPM for their dedication and
efforts in this campaign.

{\bf We thank the referee J. Osborn, as well as an anonymous
referee, for their careful reading of the manuscript which
resulted in significant improvements to this work.}

Partial support from DGAPA (Universidad Nacional Aut\'onoma de
M\'exico) PAPIIT projects IN109809, IN122409 and IT104311-2 is
gratefully acknowledged.


\bibliographystyle{mnras}
\bibliography{spmsite}


\label{lastpage}
\end{document}